\documentclass{article}

\usepackage{mlspconf} 

\usepackage[utf8]{inputenc}             
\usepackage[T1]{fontenc}                
\usepackage{url}                        
\usepackage{booktabs}                   
\usepackage{amsfonts}                   
\usepackage{microtype}                  
\usepackage{xcolor}                     
\usepackage{graphicx}                   
\usepackage{amsmath}                    
\usepackage[caption=false]{subfig}      
\usepackage{etoolbox}                   
\usepackage{tikz}                       
\usepackage{float}                      
\usepackage{adjustbox}                  
\usepackage{nicefrac}                   
\usepackage[round-mode=places,detect-weight=true,detect-inline-weight=math]{siunitx} 
\usepackage{breakcites}                 
\usepackage{array}                      %
\usepackage[breaklinks=true]{hyperref}  
\usepackage{xargs}                      %
\usepackage{comment}
\usepackage{enumitem}
\usepackage{url}
\usepackage{hyperref}

\input{utilities/underscore_replacement}
\makeatletter

\newcommand*{\@rowstyle}{}

\newcommand*{\rowstyle}[1]{
  \gdef\@rowstyle{#1}%
  \@rowstyle\ignorespaces%
}

\newcolumntype{=}{
  >{\gdef\@rowstyle{}}%
}

\newcolumntype{+}{
  >{\@rowstyle}%
}

\makeatother

\newcommandx{\doublerow}[2][2]{ #1 & #2  \\}
\newcommandx{\triplerow}[3][2,3]{ #1 & #2 & #3 \\}
\newcommandx{\quadrow}[4][2,3,4]{ #1 & #2 & #3 & #4 \\}
\newcommandx{\pentrow}[5][2,3,4,5]{ #1 & #2 & #3 & #4 & #5 \\}
\newcommandx{\hexarow}[6][2,3,4,5,6]{ #1 & #2 & #3 & #4 & #5 & #6 \\}

\newcommandx{\aefdbcrowx}[6][2,3,4,5,6]{ #1 & #6 & #4 & #2 & #3 & #5 \\}

\makeatletter 
    \def\fps@figure{hbtp}
    \def\fps@table{hbtp}
\makeatother

\newcommand{\figurewidth}{.99\linewidth}
\newcommand{\tablewidth}{.99\linewidth}
\newcommand{\subfigurewidth}{.47\linewidth}

\newcommand{\RR}{I\!\!R} 
\newcommand{\Nat}{I\!\!N} 
\newcommand{\CC}{I\!\!\!\!C} 

\DeclareUnicodeCharacter{200B}{{\hskip 0pt}}
\graphicspath{ {./figures/} }
\title{Leveraging Domain Features for Detecting Adversarial Attacks Against Deep Speech Recognition in Noise}

\name{Christian Heider Nielsen and Zheng-Hua Tan\thanks{Corresponding Author: Zheng-Hua Tan (zt@es.aau.dk).}}
\address{Department of Electronic Systems, Aalborg University, Denmark}

\begin{document}

\date{}
\maketitle


\begin{abstract}
In recent years, significant progress has been made in deep model-based automatic speech recognition (ASR), leading to its widespread deployment in the real world. At the same time, adversarial attacks against deep ASR systems are highly successful. Various methods have been proposed to defend ASR systems from these attacks. However, existing classification based methods focus on the design of deep learning models while lacking exploration of domain specific features. This work leverages filter bank-based features to better capture the characteristics of attacks for improved detection. Furthermore, the paper analyses the potentials of using speech and non-speech parts separately in detecting adversarial attacks. In the end, considering adverse environments where ASR systems may be deployed, we study the impact of acoustic noise of various types and signal-to-noise ratios. Extensive experiments show that the inverse filter bank features generally perform better in both clean and noisy environments, the detection is effective using either speech or non-speech part, and the acoustic noise can largely degrade the detection performance.
\end{abstract}



\section{Introduction}

Numerous successful adversarial attack methods on automatic speech recognition (ASR) systems have been proposed \cite{Iter2017, Carlini2018, schonherr_adversarial_2018, Alzantot2018, Cisse2017a, Kreuk2018, Qin2019a, Taori2019a}.
They demonstrate that through small optimised perturbations to an input signal, it is possible to fool an ASR system to produce an alternative targeted result, while the perturbations are largely 
imperceptible for humans. 
Prevalence of ASR is on the rise, with already rolled out systems as Google Assistant, Amazon Alexa, Samsung Bixby, Apple Siri and Microsoft Cortana. And with them comes an ever increasing attack surface.
The aforementioned systems are potentially responsible for home automation, e.g. turning on and off devices, and for administrative tasks, e.g. making purchases online. 
  Adversarial attacks can make a voice assistant behave maliciously and thus cause a significant threat to the security, privacy, and even safety of its user \cite{yan2022survey}. 
  Obviously, without validating the request being legitimate will leave security holes in users systems.

The nature of attacks can be targeted or untargeted. While the goal of untargeted attacks is to force the ASR model to produce an incorrect class, targeted attacks aim to make the model output a predetermined class and thus are most dangerous in terms of intrusive behaviour. 
In this work, we will focus on targeted attacks only.

Strategies for defending adversarial attacks can be categorised as proactive and reactive  \cite{Hu2019}. In proactive defences, one seeks to build more robust ASR models, e.g. through training with adversarial examples. With reactive approaches, one aims to detect the existence of adversarial attacks at testing time. 
There exist a number of reactive defence methods. The works in \cite{Das2019, rajaratnam_isolated_nodate, subramanian_robustness_2019, rajaratnam2018noise} defend against audio adversarial attacks by preprocessing a speech signal prior to passing it onto the ASR system. An unsupervised method with no need for labelled attacks is presented in \cite{Akinwande2020}, where the defence is realised using anomalous pattern detection. 
Rather than detecting adversarial examples, the work in \cite{yang_characterizing_2019} characterises them using temporal dependencies. 

In \cite{samizade_adversarial_2019}, adversarial example detection is formulated as a binary classification problem, in which a convolutional neural network (CNN) is used as the model and Mel-frequency cepstral coefficients (MFCCs) are used as the feature. While MFCC is the most commonly used feature for numerous applications, it remains largely uncontested for this particular defence purpose. Our preliminary study shows that attacks exhibit noticeable energy permutation in high frequency regions, where the resolution of MFCCs is by design sacrificed. 
Furthermore, it is shown in \cite{yu2017spoofing} that cepstral features from learned filter banks that have denser spacing in the high frequency region are more effective in detecting spoofing attacks. 
We are therefore motivated to contest the MFCC representation with alternative cepstral features that refocus the spectral resolution in different frequency regions. 

ASR systems when deployed in the real world will be exposed to adverse acoustic environments. However, detecting adversarial attacks in noise has rarely been investigated in the literature. In this work, we study the impact of acoustic noise on the performance of adversarial attack detection. It is relevant to mention that the success rate of attacking in noisy environments will decrease as briefly shown in \cite{Yuan2018a} and that the noise can change ASR output \cite{rajaratnam2018noise}, which is of interest for future study in a systematic way. Furthermore, it is unknown how speech and non-speech parts are useful for adversarial attack detection. This work studies their potentials.

The contribution of this paper is four-fold. 
\begin{itemize}
    \item To our knowledge, this is the first systematic study of various cepstral features for adversarial attack detection and of impact of acoustic noise, speech only, and non-speech only on detection performance. These include MFCC, inverse MFCC (IMFCC), Gammatone frequency cepstral coefficients (GFCC), inverse GFCC (IGFCC) and linear frequency cepstral coefficients (LFCC). These features are not new, but they (except for MFCC) have not been explored for adversarial attack detection. In addition, their use is highly motivated, and significant improvements are obtained using inverse filter banks as in iMFCC and iGFCC. The study of these features further helps understand the characteristics of the adversarial attacks. 
    \item This work systematically studies the impact of adverse environments for the first time and gains much insight.
    \item The paper analyses the potentials of using speech and non-speech parts separately in detecting attacks.
    \item A significant effort has been put on making the work reproducible. The source code for reproducing the results with the parameterisation used in this work and a full catalogue of results are made publicly available \footnote{\url{https://aau-es-ml.github.io/adversarial_speech_filterbank_defence/}}.

\end{itemize}

The paper is organized as follows. Section II presents background and analysis of audio adversarial attacks. Methodology and experiment design are provided in Section III. Section IV provides experimental results and discussions. The work is concluded in Section V.

\section{Background and Analysis}

This section presents both white-box and black-box attacks, analyses spectral representations of attacks, and introduces filter banks applied in this work. 

\subsection{White-box and black-box attacks}
This work considers both white-box and black-box attacks.
In white-box attacks, the adversary has full access to the parameters of the victim model, while in black-box attacks, the adversary does not have the access to the model parameters. An examples of state-of-the-art white-box attack methods against ASR systems is the gradient-based Carlini \& Wagner method \cite{Carlini2018}. An example of black-box attack methods is the gradient-free Alzantot method \cite{Alzantot2018} that uses genetic algorithm optimization. Gradient approximation is used to generate black-box audio adversarial examples in \cite{huang2022generation}. 

We build our work on the publicly-available adversarial attack data sets provided by our early work in \cite{samizade_adversarial_2019}. 
For white-box attacks, the Carlini \& Wagner method \cite{Carlini2018} is used to attack Baidu DeepSpeech model 
\cite{battenberg_exploring_2017}. The DeepSpeech model is trained using the publicly available Mozilla Common Voice data set \cite{Mozilla}. 
For black-box attacks, the Alzantot method \cite{Alzantot2018} is applied to a keyword spotting deep model \cite{sainath_convolutional_nodate} trained with the publicly available Google Speech Command data set \cite{warden2018speech}. All data sets are sampled at 16 kHz. 

\begin{figure} 
    \centering
        {
        \includegraphics[width=\subfigurewidth]{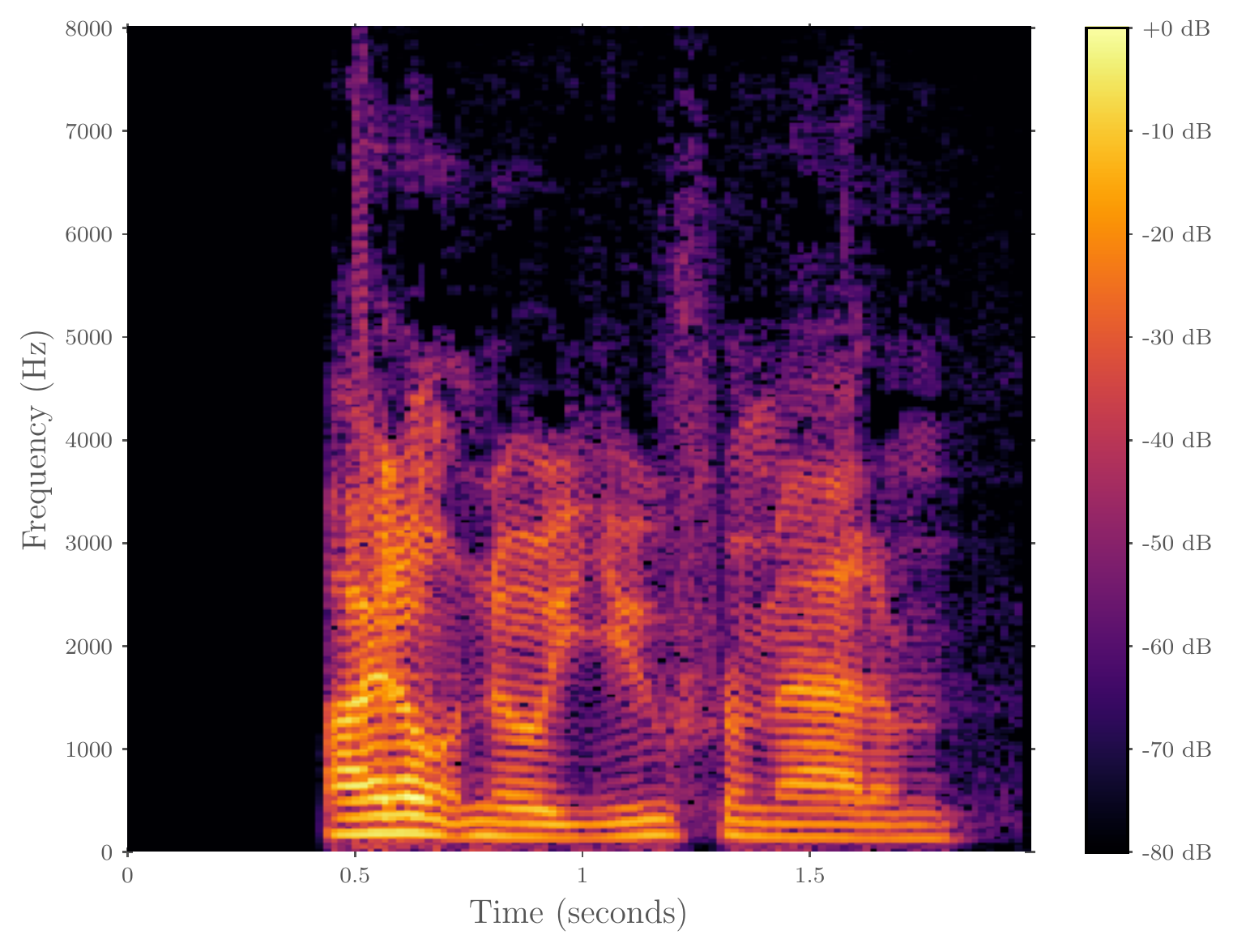}
      }
      {
                \includegraphics[width=\subfigurewidth]{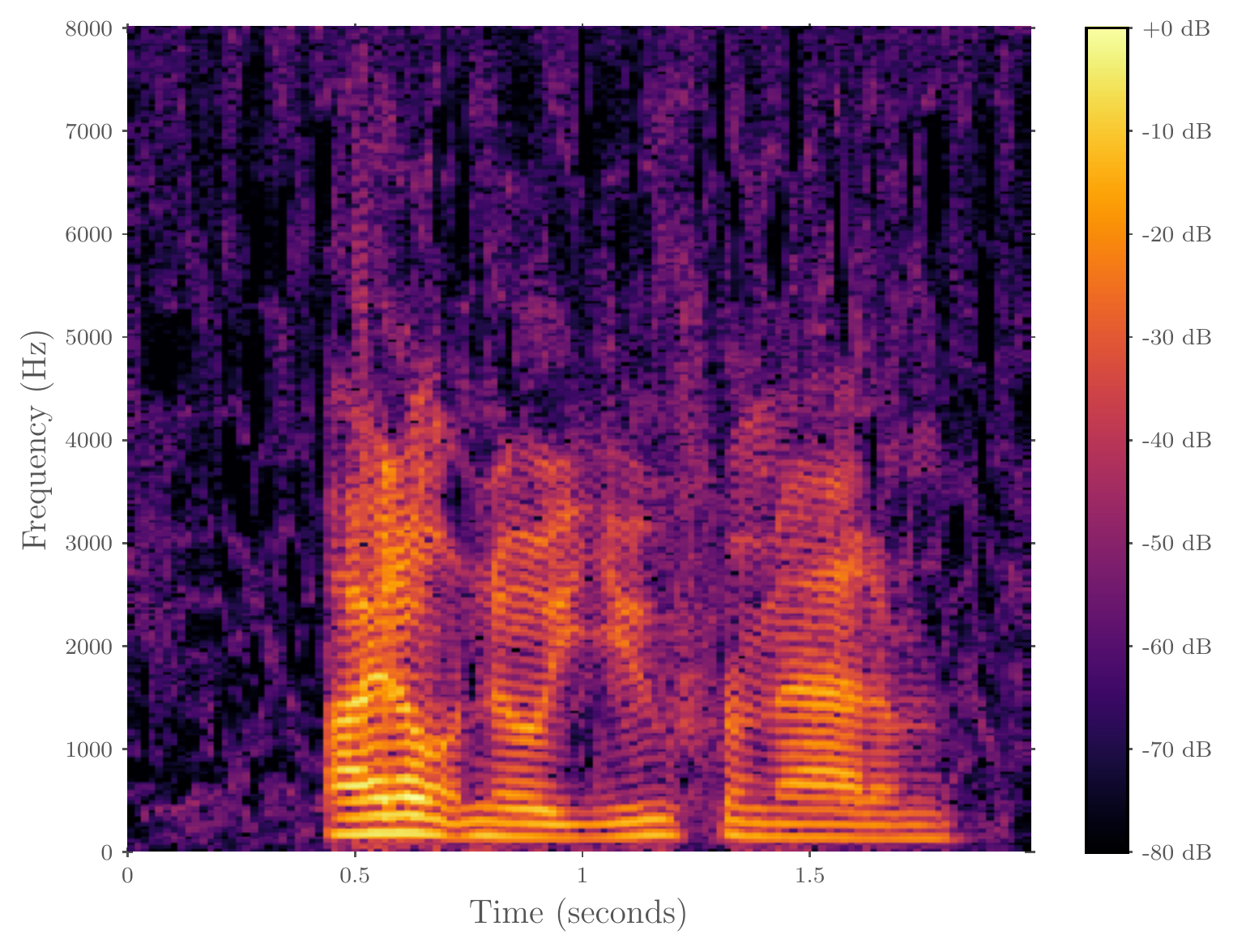}
      }
    \caption{Spectrogram of the utterance "I don't know you from Adam", without (left) and with (right) white-box adversarial attack. x-axis is time in seconds and y-axis is frequency in Hz.}\label{fig:stfts}
\end{figure} 

\subsection{Spectral analysis}

Figure \ref{fig:stfts} shows the spectrograms of both benign and attacked samples, calculated with a frame length of $32$ ms and a frame shift of $16$ ms. Inspecting the figure, we hypothesise that for detecting attacks, using filter banks that concentrate resolution in the high frequency regions is beneficial. 

\begin{figure}
    \centering
    \includegraphics[width=7.8cm]{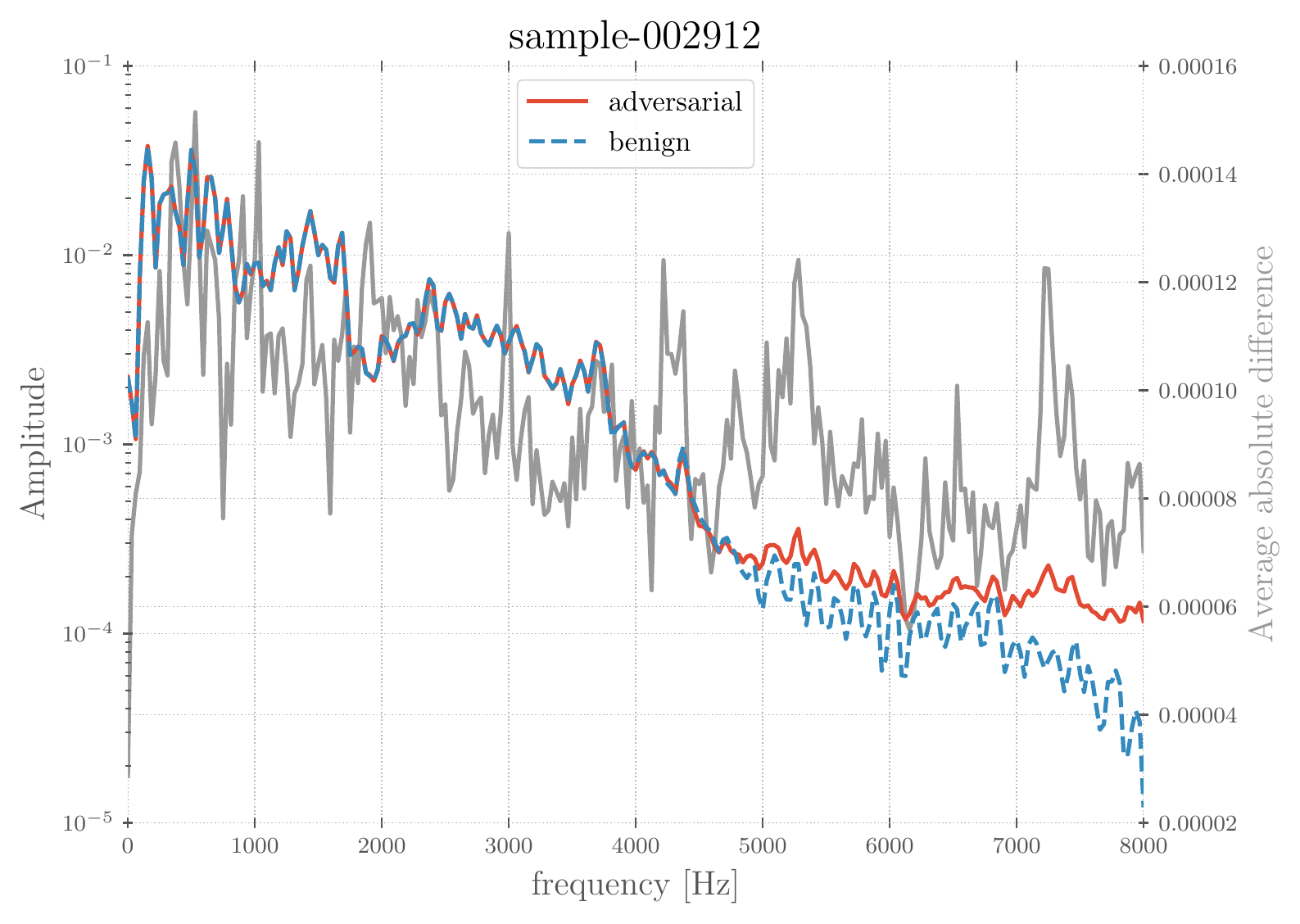}
    \caption{Long term average spectra of both benign and adversarial utterances of "I don't know you from Adam" and the average absolute frame-level difference between them.}\label{fig:spectrum_average}
\end{figure}

Figure \ref{fig:spectrum_average} shows an example of the long term average spectrum (LTAS) calculated using Welch's method \cite{welch1967use} for the benign and adversarial signals, together with the average absolute difference between the two spectra. The difference is computed by taking the spectrograms of both benign and adversarial signals from Fig. \ref{fig:stfts}, subtracting them frame-wise and averaging the absolute residuals.
Contrasting the LTAS and the average absolute difference further highlights the importance of high-frequency regions, since the speech signals have much lower energy in  high frequency regions and thus perturbations in high frequency regions become \emph{relatively} more prominent. 

\begin{figure} 
    \centering
            \subfloat{
        \includegraphics[width=\subfigurewidth]{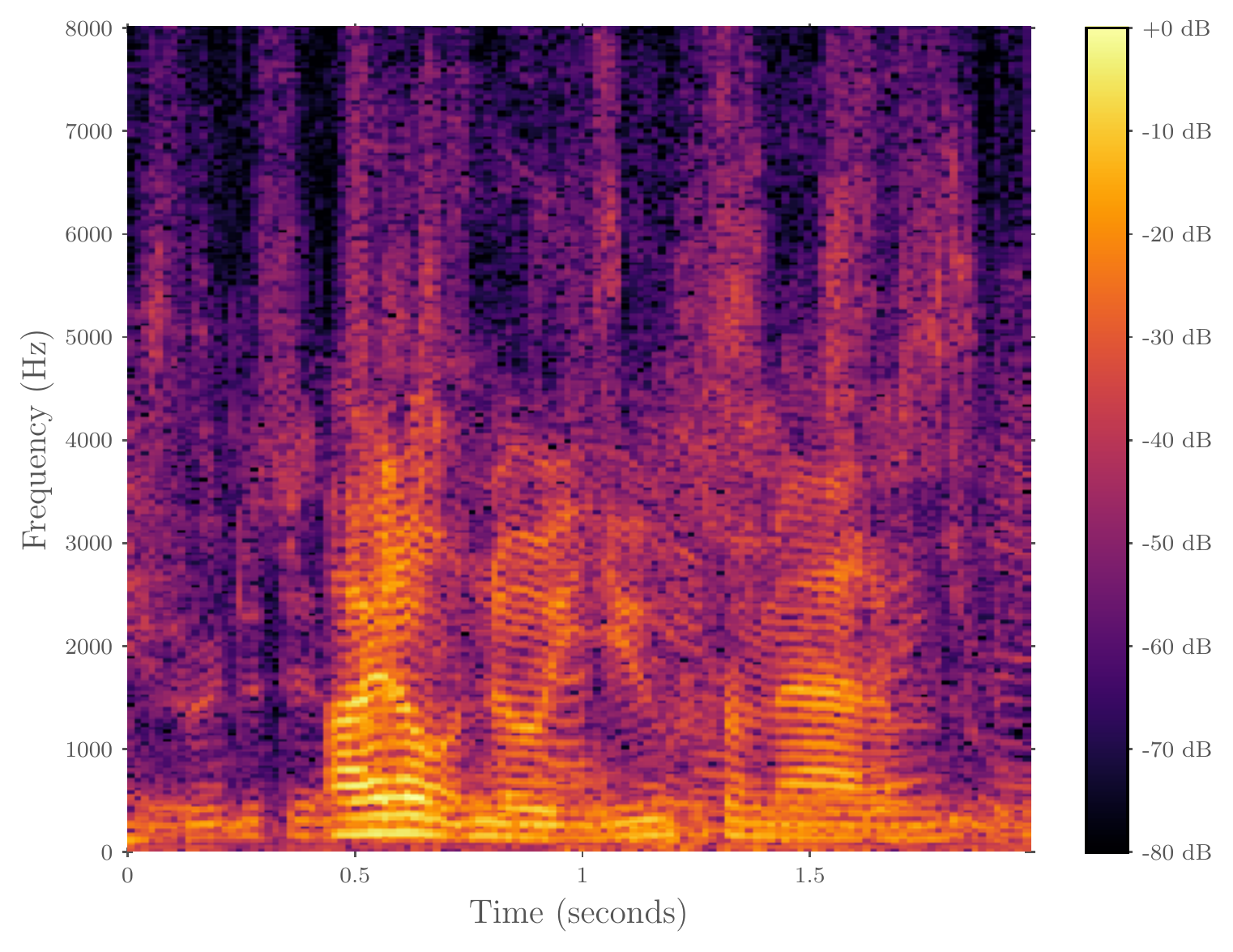}
      }
      \hfill
          \subfloat{
        \includegraphics[width=\subfigurewidth]{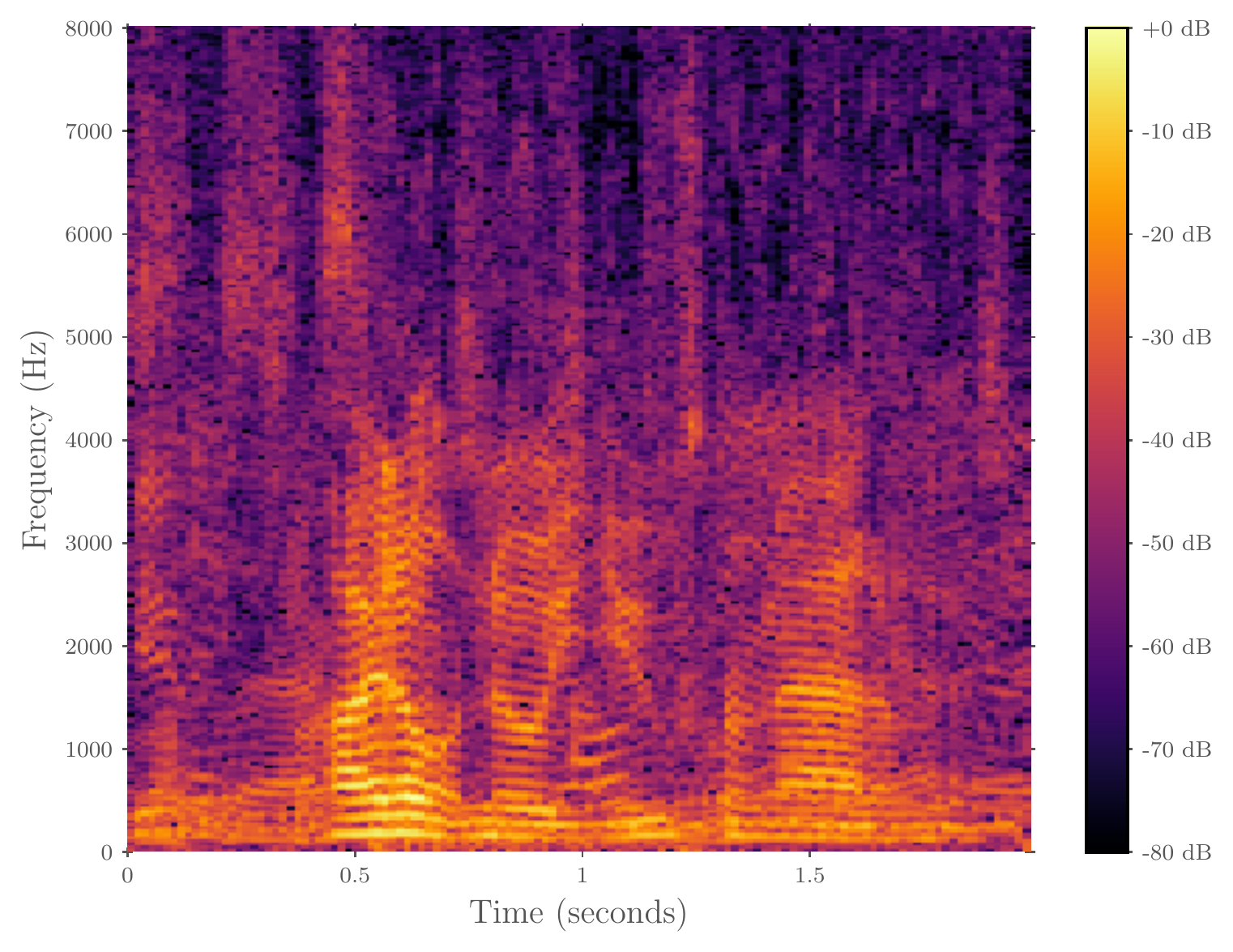}
      }
    \caption{Spectrogram of the utterance "I don't know you from Adam" mixed with babble noise at 10dB SNR, without (left) and with (right) white-box adversarial attack.}\label{fig:adv_noise_10_snr}
\end{figure}

\subsection{Noisy attack example}

Figure \ref{fig:adv_noise_10_snr} shows two spectra: one for the benign signal and the other for the corresponding attacked signal, both corrupted with babble noise at 10dB SNR. It is clear that the additive noise makes it more difficult to detect attacks. 

\subsection{Filter banks} 

Filter banks are widely used to model the frequency selectivity of an auditory system. Traditionally, the design of filter banks is motivated by psychoacoustic experiments. For example, humans are better at discriminating lower frequencies than higher frequencies, and the Mel-scale triangular filter bank exploits the characteristics of human perception of sound by having higher resolution at lower frequencies than at higher frequencies. Other filter banks are equivalent rectangular bandwidth (ERB) scale Gammatone filter bank and linear triangular filter bank. In this work we also apply the inverse variants of these filter banks.  
Figure \ref{fig:filter_banks} illustrates the regular and inverse variant of the Mel-scale triangular filter banks.

\begin{figure} 
    \centering
    \foreach \transName in {mel-scale,inverse_mel-scale} {
        \subfloat[20 channel \reprname{\transName} filter bank\label{fig:filter_bank_\transName}]{
            \includegraphics[width=6cm]{colorful/filterbanks/\transName}
      }
      \hfill 
      \hspace{4cm}
    }
    \caption{Mel-scale triangular filter bank and its inverse. x-axis is frequency in Hz and y-axis is gain.}\label{fig:filter_banks}
\end{figure}

\subsubsection{Linear triangular filter bank}
In a linear triangular filter bank, the gain $H_m(k)$ of the $k'th$ data point of the $m'th$ triangular band filter is as follows \cite{10.5555/560905}

\begin{equation}\label{equ:tri}
\resizebox{\linewidth}{!}{%
    $H_m(k) = \begin{cases}
            \frac{k - f_b(m - 1)}{f_b(m) - f_b(m - 1)} & f_b(m-1) \le k \le f(m) \\
            \frac{f_b(m + 1) - k}{f_b(m + 1) - f_b(m)} & f_b(m) \le k \le f_b(m+1) \\
            \hfil 0 \hfil                                                             & {\text{else}} \\ 
            \end{cases}$
            }
\end{equation}

\noindent where $f_b(m)$ is the boundary frequency point of the $m'th$ filter

\begin{equation}\label{equ:li}
f_b(m) = f_{low} + m\frac{f_{high} - f_{low}}{M+1}
\end{equation}

\noindent where 
$M$ is the number of filters, $f_{low}$ and $f_{high}$ are the frequency range $[0,\frac{F_s}{2}]$ and $F_s$ is the sampling frequency in Hz.

\subsubsection{Mel-scale triangular filter bank}

For a Mel-scale triangular filter bank, we construct a triangular filter bank with the boundary frequencies $f_b(m)$ equally spaced on the Mel-scale 

\begin{equation}\label{equ:Mel}
\resizebox{\linewidth}{!}{%
$
f_b(m) = Mel^{-1}(Mel(f_{low}) + m\frac{Mel(f_{high}) - Mel(f_{low})}{M+1})
$}
\end{equation}

\noindent where
$Mel(f) = 1125 ln(1 + (f/700)$ and
$Mel^{-1}(m) = 700 (e^{m/1125} - 1)$.

\subsubsection{Gammatone filter bank}

For a Gammatone filter bank the center frequencies $f_c(g)$ are equally spaced on the equivalent rectangular bandwidth (ERB) scale $ERB(f) = \frac{f}{9.26} + 24.7$. 


Then the gain $H_g(k)$ of the $k'th$ data point of the $g'th$ the gammatone wavelet filter is computed as
\begin{equation}
H_g(k) = \frac{(L - 1)!}{ (ERB(f_c(g))  + j(2 \pi k - 2 \pi f_c(g) ))^L}
\end{equation}

\noindent where $L$ is the filter order, typically set to $L=4$, and $f_c(g)$ is the center frequency of the $g'th$ filter

\begin{equation}
f_c(g) = -C + (f_{high} + C) e^{g * log_{10}(\frac{f_{low}+C}{f_{high} + C}) / G}
\end{equation}

\noindent where $C=228.83$, $1 \leq g \leq G$ and $G$ is the number of filters.

\subsubsection{Inverse filter banks}

Producing the inverse filter bank variant is as simple as reversing the regular filter bank with respect to the frequencies.

\section{Methodology and experiment design}

\subsection{Data sets}
The adversarial attack data sets used in this work are from \cite{samizade_adversarial_2019}, which are publicly available \footnote{\url{https://github.com/zhenghuatan/Audio-adversarial-examples}}. We will refer to this article for the specifics on the generation of the adversarial attacks. 
Data set A (white-box) and data set B (black-box) have $1470$ and $2660$ samples, respectively. Data set A contains $620$ adversarial samples and $850$ benign ones, and Dataset B contains $1252$ adversarial samples and $1408$ benign ones, which are reasonably balanced data sets. We conduct extensive experiments using data set A for training and data set B for testing, and vice versa, to investigate the robustness of the proposed methods across types of adversarial attacks. 

The data sets provided by \cite{samizade_adversarial_2019} comprise utterances of varying lengths. In order to use a model with inputs of fixed size,
we chop the utterances into blocks of $512$ ms with a shift of $512$ ms, i.e. no overlap. This is in contrast to \cite{samizade_adversarial_2019}, where zero-padding is applied to match the longest utterance in the data sets. The issues with zero-padding are that the length of the longest utterance in the data set (rather than the problem domain itself) determines the model size, and computational resources are wasted by adding many zeros. Utterance-level decision can be made easily by later fusion of block level decisions, which is out of the scope of this work. 

After block chopping, we split the data into training, validation and test subsets. In order to make the data split unbiased and fully reproducible, we use a hash function-based selection on the utterance filenames, targeting at $70\%$ for training, $10\%$ for validation and $20\%$ for test. In the end, the blocks amount to $(1187,183,272)$ adversarial attack blocks and $(1550,222,377)$ benign blocks for training, validation, testing splits from data set A. From data set B we end up with $(1260,180,360)$ adversarial attack blocks and $(1398,200,400)$ benign blocks.

\subsection{Detection model}
We use CNNs to classify an input as either being benign or adversarial. The chosen model architecture and hyperparameterisation mostly follows \cite{samizade_adversarial_2019}. One change is to have a fixed input size as $(31, 20)$, corresponding to (\#frames, \#cepstral coefficients), throughout all experiments. 
Additionally our model has one single output for binary classification. This slightly modified architecture is found in Table \ref{tab:model_arch}. 
\begin{table}
\centering
\caption{Model Architecture}
\label{tab:model_arch}
\medskip
\begin{adjustbox}{width=.6\linewidth, center=.6\linewidth}
\begin{tabular}{lccccr}
\toprule
\pentrow{type}[size][field][stride][activation]
\midrule
\pentrow{Conv2d}[$64$][$(2, 2)$][$(1, 1)$][$ReLU$]
\pentrow{MaxPool2d}[$(1, 3)$][$(1, 3)$]
\pentrow{Conv2d}[$64$][$(2, 2)$][$(1, 1)$][$ReLU$]
\pentrow{MaxPool2d}[$(1, 1)$][$(1, 1)$]
\pentrow{Conv2d}[$32$][$(2, 2)$][$(1, 1)$][$ReLU$]
\pentrow{MaxPool2d}[$(2, 2)$][$(2, 2)$]
\pentrow{Flatten}[$896$]
\pentrow{Fully Connected}[$128$][][] [$ReLU$]
\pentrow{Fully Connected}[$1$][][] [$Sigmoid$]
\bottomrule
\end{tabular}
\end{adjustbox}
\end{table}

\subsection{Detection with various cepstral features}
We study a set of filter banks for attack detection aiming at investigating the importance of different frequency regions. 
In extracting speech features, each block is first segmented into frames using a frame length of $32$ ms with a frame shift of $16$ ms. Each frame is then processed by taking the discrete Fourier transform, calculating power spectral estimate, applying a filter bank, taking the logarithm of the filter bank energies and finally taking discrete cosine transform, which gives cepstral feature for the frame. This leads to $20$ cepstral coefficients for each of the $31$ frames in a block. A supervector is a concatenation of the cepstral feature vectors of a multi-frame block.


In this work we consider the following cepstral features depending on different filter banks: Mel-frequency cepstral coefficients, inverse MFCC, Gammatone frequency cepstral coefficients, inverse GFCC, and linear frequency cepstral coefficients.

\subsection{Detection using speech and non-speech parts}
Attacks may occur in both speech and non-speech regions. To investigate contributions of speech and non-speech parts to attack detection, we split the signals of the white-box data set using the open-source, unsupervised robust voice activity detection (rVAD) method \cite{tan_rvad_2020} \footnote{\url{https://github.com/zhenghuatan/rVAD}}, with the default parameterisation of their open source implementation. 
The extracted speech segments and non-speech segments from a single signal are separately concatenated into one signal for speech and another for non-speech, which each is subsequently chopped into blocks, framed and the cepstrums are computed.

\subsection{Detection in noisy environments}

\begin{figure}
    \centering
    \includegraphics[trim=0 15 0 0, clip,width=7cm]{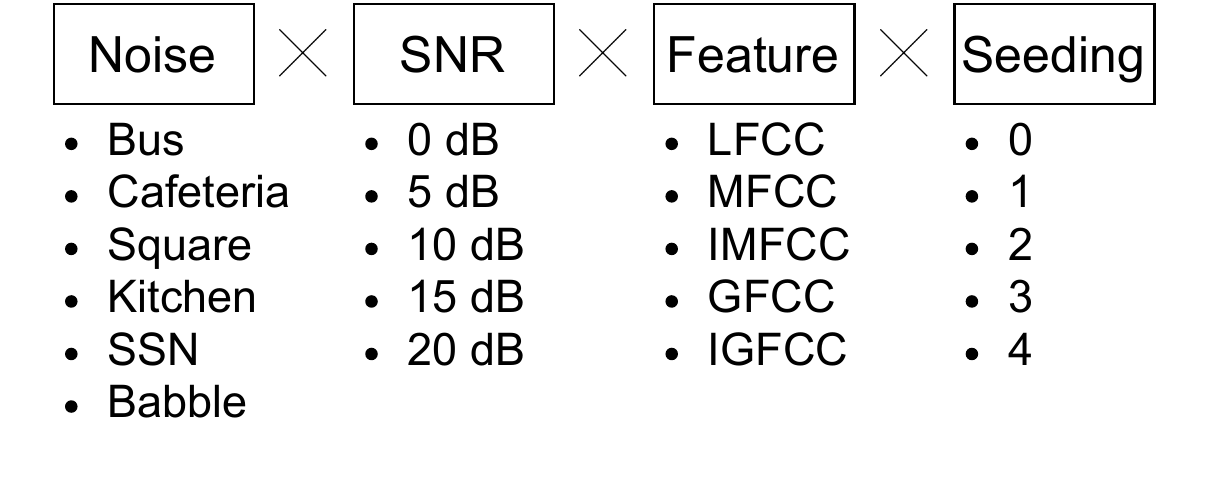}
    \caption{Combinations of models to be trained and evaluated}\label{fig:singular_noise_model_combinations}
\end{figure}

To study the impact of noise on detection performance, we consider the various filter bank models for a range of noise types and SNRs. Figure \ref{fig:singular_noise_model_combinations} shows the combinations of conditions/models in this experiment, leading to quite a number of experiments of different models. The considered noise types are cafeteria noise \textit{PCAFETER}, bus passenger noise \textit{TBUS}, city square noise \textit{SPSQUARE}, kitchen noise \textit{DKITCHEN}, speech shaped noise \textit{ssn} and babble noise \textit{bbl}.
The first four are from \cite{thiemann} and the last two from \cite{7846281}. We chose these to cover possible real world settings where ASR systems are employed and may be exposed to adversarial attacks, and these noise files are readily available online, making the experiments reproducible. We mix noise with speech at $[0,5,10,15,20]$ dB SNR levels.


We again split a signal into speech and non-speech parts using rVAD \cite{tan_rvad_2020}. However, this splitting is only for calculating root mean square (RMS) of the speech part. Prior to mixing we scale the noise by the RMS ratio between the speech part of the signal and the noise. We additively mix the scaled noise with the signal, at a target SNR calculated using the speech part. Prior to mixing, noise signals of each noise type are also split into training, validation and test subsets to ensure no leakage among them. 

\section{Experimental results} 

In this section, we present the experimental results. For performance evaluation, we use the receiver operator characteristic's area under curve (ROCAUC) metric. In all result tables we will display the mean and standard deviation of $5$ independent runs of varying seeds for each experiment. The seeding affects the model parameter initialisation and batch sampling order of the same fixed training, validation and test splits. 

\subsection{Cepstral features for attack detection}\label{exp_setup:sep_white_black}

This experiment evaluates the five cepstral features on different white- and black-box attack combinations. To make the results comparable across attack types, we truncate the number of examples in each dataset to be the same as the lesser. 

The results are shown in Table \ref{tab:rocauc_ab}. First  we observe that inverse filter bank features IGFCC and IMFCC perform the best, GFCC and MFCC are inferior, and LFCC is in between. Next, we can see the generalisation performance across data sets with different training and test combinations. Training with black-box and testing with white-box sees a significant performance drop. It is interesting to note that IMFCC and IGFCC perform so well when training on white-box and testing on black-box, but  the same is not observed when training on black-box and testing on white-box. 
Training and testing on black-box shows higher ROCAUC scores than training and testing on white-box, which could be because black-box attacks are more audible and easier to detect than white-box ones. Finally multi-style training combining white-box and black-box helps.



\begin{table*}[hbt!]
\centering
\caption{Mean and standard deviation of ROCAUC for scenarios where both training and test tests containing white-box and black-box examples. \textit{w\&b} denotes both white-box and black-box attacks.}
\label{tab:rocauc_ab}
\medskip
\begin{adjustbox}{width=12cm, center=\tablewidth}
\begin{tabular}{lccccc}
\toprule
\aefdbcrowx{filter bank}                           [GFCC]             [IGFCC]                      [IMFCC]                     [LFCC]              [MFCC]
\aefdbcrowx{train set \hspace{0.5cm} test set}
\midrule
\aefdbcrowx{white-box}
\aefdbcrowx{\hspace{1.5cm} white-box}         [$.915 \pm .02$ ][$\mathbf{.981 \pm .003}$] [$.98 \pm .003$]          [$.961 \pm .003$][$.941 \pm .013$]
\aefdbcrowx{\hspace{1.5cm} black-box}         [$.694 \pm .04$ ][$.932 \pm .012$]          [$\mathbf{.946 \pm .01}$] [$.692 \pm .127$][$.576 \pm .072$]  
\aefdbcrowx{black-box}
\aefdbcrowx{\hspace{1.5cm} white-box}         [$.528 \pm .011$][$.676 \pm .034$]          [$\mathbf{.672 \pm .091}$][$.625 \pm .057$][$.527 \pm .015$] 
\aefdbcrowx{\hspace{1.5cm} black-box}         [$.982 \pm .004$][$\mathbf{.991 \pm .005}$] [$.991 \pm .007$]         [$.988 \pm .003$][$.983 \pm .005$]            
\aefdbcrowx{w\&b-box}
\aefdbcrowx{\hspace{1.5cm} w\&b-box}   [$.948 \pm .006$][$.986 \pm .001$]          [$\mathbf{.989 \pm .002}$][$.979 \pm .001$][$.966 \pm .001$]
\aefdbcrowx{\hspace{1.5cm} white-box}         [$.942 \pm .007$][$.985 \pm .002$]          [$\mathbf{.989 \pm .002}$][$.976 \pm .001$][$.964 \pm .002$]
\aefdbcrowx{\hspace{1.5cm} black-box}         [$.972 \pm .012$][$\mathbf{.99 \pm .005}$]  [$.99 \pm .007$]          [$.988 \pm .003$][$.974 \pm .004$]
\bottomrule
\end{tabular}
\end{adjustbox}

\end{table*}

\bigskip

\begin{table*}[hbt!]

\centering
\caption{Mean and standard deviation of ROCAUC for speech and non-speech data sets of white-box attacks. \textit{speech \& ns} denotes both speech and non-speech attack datasets.}
\label{tab:rocauc_speechsilence}
\medskip
\begin{adjustbox}{width=12cm, center=\tablewidth}
\begin{tabular}{lccccc}
\toprule
\aefdbcrowx{filter bank}                       [GFCC]              [IGFCC]                     [IMFCC]                     [LFCC]             [MFCC]
\aefdbcrowx{train set \hspace{0.5cm} test set}
\midrule                          
\aefdbcrowx{non-speech}
\aefdbcrowx{\hspace{1.5cm} non-speech}       [$.965 \pm .014$] [$\mathbf{.995 \pm .004}$][$.993 \pm .006$]         [$.988 \pm .006$][ $.97 \pm .016$]
\aefdbcrowx{\hspace{1.5cm} speech}        [$.661 \pm .058$] [$\mathbf{.851 \pm .071}$][ $.823 \pm .06$]         [$.746 \pm .042$][$.724 \pm .074$]
\aefdbcrowx{speech }
\aefdbcrowx{\hspace{1.5cm} speech}        [$.679 \pm .036$] [$\mathbf{.945 \pm .018}$][$.944 \pm .021$]         [$.886 \pm .028$][$.833 \pm .051$]
\aefdbcrowx{\hspace{1.5cm} non-speech}       [$.863 \pm .033$] [$\mathbf{.988 \pm .011}$][$.987 \pm .009$]         [ $.948 \pm .02$][$.937 \pm .017$] 
\aefdbcrowx{speech \& ns}
\aefdbcrowx{\hspace{1.5cm} s\&ns} [$.894 \pm .006$] [$.978 \pm .005$]         [$\mathbf{.982 \pm .008}$][$.941 \pm .013$][$.929 \pm .013$]
\aefdbcrowx{\hspace{1.5cm} non-speech}       [$.973 \pm .01$]  [$.997 \pm .001$]         [$\mathbf{.999 \pm .002}$][ $.98 \pm .016$][$.986 \pm .002$]
\aefdbcrowx{\hspace{1.5cm} speech}        [$.858 \pm .008$] [ $.97 \pm .007$]         [$\mathbf{.975 \pm .011}$][$.925 \pm .015$][$.903 \pm .019$]
\bottomrule
\end{tabular}
\end{adjustbox}

\end{table*}

\subsection{Attack detection in speech and in non-speech}\label{exp_setup:sep_silence_speech}
This experiment compares performance in both speech and non-speech regions as well as across them to see if a model trained for speech only is able to generalise to non-speech only and vice-versa. To make the results comparable we truncate the size of the speech and non-speech data sets to make them match. For splitting the sample into the speech and non-speech parts, we apply the rVAD method \cite{tan_rvad_2020}, as detailed earlier. 

Table \ref{tab:rocauc_speechsilence} shows the five cepstral features on speech and non-speech parts of white-box attack with different training and test set combinations. From the results we observe that adversarial attacks can be detected via both speech and non-speech regions effectively and that inverse filter bank features remain performing the best. It is interesting to note that testing on non-speech always performs better than testing on speech, no matter the model is trained in non-speech, speech or a mixture of speech and non-speech, which indicates non-speech part contains a stronger cue of adversarial attacks. Training on both speech and non-speech and testing on non-speech gives the highest ROCAUC scores for almost all cepstral features.  

\subsection{Attack detection in noise}\label{exp_setup:sep_noise}

In this experiment we evaluate the various filter bank models for a range of noise types and SNRs for white-box attacks.

\subsubsection{Noise and SNR specific training}
Table \ref{tab:rocauc_noisetypes} presents results for the narrowly trained models using five features for differing noise types and SNRs. We observe a general tendency that using inverse filter bank features outperforms the rest with the only exception being the \textit{kitchen} noise. This could be explained by the fact that the \textit{kitchen} noise has much flatter long term amplitude spectrum, as we found experimentally, which leads to LFCC performing the best. Across all noise types and SNRs, IGFCC performs the best, for which the reason could be that Gammatone filter banks are known to be noise robust. In all cases, noise significantly decreases the detection performance, and in some 0dB cases, the performance is close to random guess. 

\begin{table*}[ht!]
\centering
\caption{Mean and standard deviation of ROCAUC for  data sets of various noise types and various SNR levels using five different cepstral coefficients. White-box attack data set is used. \textit{avg} is calculated across all seedings (of five independent runs) and SNRs, \textit{all} denotes all noise types and \textit{avg\_all} is the average over all noise types and SNRs. Noise type and SNR are matched for both training and testing in all settings, namely a narrowly trained systems is evaluated in this table.
}
\bigskip

\label{tab:rocauc_noisetypes}
\medskip
\begin{adjustbox}{width=12cm, center=\tablewidth}
\begin{tabular}{lccccc}
\toprule
\aefdbcrowx{cepstral features}[GFCC][IGFCC][IMFCC][LFCC][MFCC]
\aefdbcrowx{noise type \hspace{0.2cm} snr}
\midrule
\aefdbcrowx{bbl}
\aefdbcrowx{\hspace{1.5cm} 0db}[$.528 \pm .029$][$\mathbf{.687 \pm .059}$][$.654 \pm .027$ ][$.573 \pm .046$ ][$.533 \pm .047$]
\aefdbcrowx{\hspace{1.5cm} 5db}[$.576 \pm .02$][$\mathbf{.806 \pm .048}$][$.793 \pm .032$ ][$.702 \pm .022$ ][$.621 \pm .042$]
\aefdbcrowx{\hspace{1.5cm} 10db }[$.668 \pm .04$][$\mathbf{.901 \pm .011}$][$.878 \pm .016$ ][$.78 \pm .02$ ][$.706 \pm .038$]
\aefdbcrowx{\hspace{1.5cm} 15db }[$.742 \pm .025$ ][$\mathbf{.951 \pm .015}$][$.921 \pm .024$ ][$.857 \pm .012$ ][$.798 \pm .017$]
\aefdbcrowx{\hspace{1.5cm} 20db }[$.838 \pm .033$ ][$\mathbf{.955 \pm .009}$][$.953 \pm .009$ ][$.901 \pm .022$ ][ $.867 \pm .03$]
\aefdbcrowx{\hspace{1.5cm} \textit{avg}}[$.671 \pm .118$ ][$\mathbf{.86 \pm .109}$][ $.84 \pm .111$ ][$.763 \pm .121$ ][$.705 \pm .127$]
\aefdbcrowx{ssn}
\aefdbcrowx{\hspace{1.5cm} 0db}[ $.506 \pm .012$ ][$\mathbf{.747 \pm .018}$ ][$.533 \pm .045$ ][ $.551 \pm .03$ ][ $.54 \pm .037$]
\aefdbcrowx{\hspace{1.5cm} 5db}[ $.518 \pm .029$ ][$\mathbf{.862 \pm .033}$ ][$.611 \pm .023$ ][$.541 \pm .017$ ][$.554 \pm .032$]
\aefdbcrowx{\hspace{1.5cm} 10db }[ $.546 \pm .049$ ][$\mathbf{.926 \pm .015}$ ][$.658 \pm .038$ ][$.69 \pm .05$ ][$.592 \pm .034$]
\aefdbcrowx{\hspace{1.5cm} 15db }[ $.691 \pm .016$ ][$\mathbf{.947 \pm .003}$ ][$.868 \pm .027$ ][$.802 \pm .037$ ][ $.73 \pm .037$]
\aefdbcrowx{\hspace{1.5cm} 20db }[ $.757 \pm .035$ ][$\mathbf{.968 \pm .006}$ ][$.937 \pm .011$ ][$.908 \pm .024$ ][$.824 \pm .045$]
\aefdbcrowx{\hspace{1.5cm} \textit{avg}}[ $.603 \pm .107$ ][$\mathbf{.89 \pm .083}$][ $.721 \pm .16$ ][$.699 \pm .148$ ][$.648 \pm .118$]
\aefdbcrowx{kitchen}
\aefdbcrowx{\hspace{1.5cm} 0db}[ $\mathbf{.562 \pm .023}$ ][$.502 \pm .005$ ][ $.511 \pm .02$ ][$.533 \pm .023$ ][$.541 \pm .039$]
\aefdbcrowx{\hspace{1.5cm} 5db}[ $.592 \pm .018$ ][ $.548 \pm .03$ ][$.536 \pm .022$ ][$.593 \pm .007$ ][$\mathbf{.6 \pm .017}$]
\aefdbcrowx{\hspace{1.5cm} 10db }[ $.652 \pm .032$ ][$.563 \pm .038$ ][$.562 \pm .013$ ][$\mathbf{.7 \pm .022}$ ][$.692 \pm .023$]
\aefdbcrowx{\hspace{1.5cm} 15db }[$.738 \pm .04$ ][$.653 \pm .032$ ][$.647 \pm .025$ ][$\mathbf{.808 \pm .035}$ ][$.799 \pm .048$]
\aefdbcrowx{\hspace{1.5cm} 20db }[ $.832 \pm .055$ ][ $.806 \pm .05$ ][$.791 \pm .032$ ][$\mathbf{.876 \pm .035}$ ][$.858 \pm .013$]
\aefdbcrowx{\hspace{1.5cm} \textit{avg}}[$.675 \pm .106$ ][$.615 \pm .114$ ][$.609 \pm .106$ ][$\mathbf{.702 \pm .133}$ ][$.698 \pm .124$]
\aefdbcrowx{cafeteria}
\aefdbcrowx{\hspace{1.5cm} 0db}[ $.508 \pm .011$ ][ $.513 \pm .022$ ][ $\mathbf{.55 \pm .045}$ ][$.532 \pm .035$ ][$.523 \pm .027$]
\aefdbcrowx{\hspace{1.5cm} 5db}[ $.521 \pm .022$ ][ $.574 \pm .053$ ][$\mathbf{.608 \pm .053}$ ][$.581 \pm .065$ ][$.539 \pm .025$]
\aefdbcrowx{\hspace{1.5cm} 10db }[ $.593 \pm .052$ ][ $\mathbf{.735 \pm .019}$ ][$.705 \pm .025$ ][$.71 \pm .04$ ][ $.605 \pm .04$]
\aefdbcrowx{\hspace{1.5cm} 15db }[ $.662 \pm .041$ ][ $\mathbf{.829 \pm .04}$][$.808 \pm .015$ ][$.805 \pm .023$ ][ $.72 \pm .021$]
\aefdbcrowx{\hspace{1.5cm} 20db }[ $.816 \pm .02$][ $\mathbf{.9 \pm .019}$ ][$.888 \pm .025$ ][$.866 \pm .019$ ][$.814 \pm .023$]
\aefdbcrowx{\hspace{1.5cm} \textit{avg}}[ $.62 \pm .119$][ $.71 \pm .153$][$\mathbf{.712 \pm .131}$ ][$.699 \pm .135$ ][ $.64 \pm .116$]
\aefdbcrowx{square}
\aefdbcrowx{\hspace{1.5cm} 0db}[ $.521 \pm .059$ ][$\mathbf{.682 \pm .104}$ ][ $.647 \pm .115$ ][ $.637 \pm .09$ ][$.539 \pm .066$]
\aefdbcrowx{\hspace{1.5cm} 5db}[ $.607 \pm .079$ ][$\mathbf{.843 \pm .023}$ ][ $.838 \pm .009$ ][ $.784 \pm .05$ ][$.625 \pm .083$]
\aefdbcrowx{\hspace{1.5cm} 10db }[ $.763 \pm .044$ ][$.906 \pm .008$ ][ $\mathbf{.91 \pm .02}$ ][ $.877 \pm .02$ ][ $.732 \pm .04$]
\aefdbcrowx{\hspace{1.5cm} 15db }[ $.831 \pm .042$ ][$\mathbf{.946 \pm .014}$ ][ $.935 \pm .026$ ][ $.945 \pm .016$ ][ $.861 \pm .02$]
\aefdbcrowx{\hspace{1.5cm} 20db }[ $.852 \pm .023$ ][$.969 \pm .01$][ $\mathbf{.97 \pm .021}$][ $.969 \pm .009$ ][$.934 \pm .022$]
\aefdbcrowx{\hspace{1.5cm} \textit{avg}}[ $.715 \pm .141$ ][$\mathbf{.869 \pm .114}$ ][ $.86 \pm .127$][ $.843 \pm .131$ ][$.738 \pm .156$]
\aefdbcrowx{bus}
\aefdbcrowx{\hspace{1.5cm} 0db}[$.695 \pm .046$ ][$\mathbf{.816 \pm .045}$ ][$.807 \pm .038$][$.784 \pm .045$ ][$.764 \pm .042$]
\aefdbcrowx{\hspace{1.5cm} 5db}[$.798 \pm .016$ ][$.928 \pm .02$ ][$\mathbf{.928 \pm .016}$ ][ $.91 \pm .027$ ][$.865 \pm .054$]
\aefdbcrowx{\hspace{1.5cm} 10db}[$.866 \pm .019$ ][$.964 \pm .012$][$\mathbf{.98 \pm .012}$][$.953 \pm .011$ ][$.917 \pm .025$]
\aefdbcrowx{\hspace{1.5cm} 15db}[$.922 \pm .016$ ][$.973 \pm .007$][$\mathbf{.984 \pm .013}$ ][$.973 \pm .014$ ][ $.95 \pm .007$]
\aefdbcrowx{\hspace{1.5cm} 20db}[$.926 \pm .023$ ][$\mathbf{.987 \pm .008}$ ][$.986 \pm .016$][ $.974 \pm .02$ ][$.933 \pm .013$]
\aefdbcrowx{\hspace{1.5cm} \textit{avg}}[$.841 \pm .092$ ][$.933 \pm .066$][$\mathbf{.937 \pm .072}$ ][$.919 \pm .077$ ][$.886 \pm .075$]
\aefdbcrowx{\textit{all}}
\aefdbcrowx{\hspace{1.5cm} \textit{avg\_all}}[$.688 \pm .137$][$\mathbf{.813 \pm .156}$][$.78 \pm .162$][$.771 \pm .15$][$.719 \pm .145$]
\bottomrule
\end{tabular}
\end{adjustbox}
\end{table*}

\newpage

\begin{table*}[ht]
\centering
\caption{Mean and standard deviation of ROCAUC for testing on unknown noise data sets. \textit{clean} is a training set consisting clean white-box attacks, and \textit{all} is all noise types and SNRs}
\label{tab:rocauc_generalisation}
\medskip
\begin{adjustbox}{width=12cm, center=\tablewidth}
\begin{tabular}{lccccc}
\toprule
\aefdbcrowx{filter bank}[GFCC][IGFCC][IMFCC][LFCC][MFCC]
\aefdbcrowx{train set \hspace{0.5cm} test set, snr}
\midrule 
\aefdbcrowx{bbl\_all\_snr}
\aefdbcrowx{\hspace{1.5cm} rest, \hspace{0.3cm} 0db} [$.516 \pm .02$][$\mathbf{.629 \pm .009}$][$.59 \pm .02$][$.602 \pm .018$][$.553 \pm .031$]
\aefdbcrowx{\hspace{2.5cm} 5db} [$.523 \pm .031$][$\mathbf{.684 \pm .014}$][$.639 \pm .03$][$.643 \pm .02$][$.584 \pm .045$]
\aefdbcrowx{\hspace{2.5cm} 10db}[$.536 \pm .048$][$\mathbf{.746 \pm .013}$][$.703 \pm .037$][$.691 \pm .024$][$.621 \pm .064$]
\aefdbcrowx{\hspace{2.5cm} 15db}[$.548 \pm .066$][$\mathbf{.813 \pm .007}$][$.77 \pm .033$][$.748 \pm .031$][$.66 \pm .086$]
\aefdbcrowx{\hspace{2.5cm} 20db}[$.557 \pm .076$][$\mathbf{.861 \pm .01}$][$.821 \pm .024$][$.807 \pm .031$][$.698 \pm .104$]
\aefdbcrowx{\hspace{2.5cm} \textit{avg}} [$.536 \pm .055$][$\mathbf{.747 \pm .085}$][$.705 \pm .089$][$.698 \pm .077$][$.623 \pm .088$]


\aefdbcrowx{rest\_all\_snr}
\aefdbcrowx{\hspace{1.5cm} bbl, \hspace{0.3cm} 0db }[$.568 \pm .01$][$.63 \pm .01$][$\mathbf{.635 \pm .019}$][$.576 \pm .013$][$.56 \pm .019$]
\aefdbcrowx{\hspace{2.5cm} 5db}[$.615 \pm .029$][$\mathbf{.744 \pm .012}$][$.727 \pm .018$][$.609 \pm .026$][$.605 \pm .015$]
\aefdbcrowx{\hspace{2.5cm} 10db }[$.669 \pm .023$][$\mathbf{.846 \pm .015}$][$.829 \pm .017$][$.697 \pm .03$][$.673 \pm .025$]
\aefdbcrowx{\hspace{2.5cm} 15db}[$.765 \pm .033$][$\mathbf{.923 \pm .016}$][$.895 \pm .023$][$.804 \pm .041$][$.765 \pm .028$]
\aefdbcrowx{\hspace{2.5cm} 20db }[$.844 \pm .023$][$\mathbf{.954 \pm .009}$][$.94 \pm .017$][$.882 \pm .028$][$.861 \pm .025$]
\aefdbcrowx{\hspace{2.5cm} \textit{avg} }[$.692 \pm .103$][$\mathbf{.819 \pm .12}$][$.805 \pm .113$][$.713 \pm .119$][$.693 \pm .111$]

\aefdbcrowx{clean}
\aefdbcrowx{\hspace{1.5cm} all, \hspace{0.3cm}  all }[$.682 \pm .118$][$.741 \pm .156$][$.719 \pm .159$][$\mathbf{.753 \pm .149}$][$.732 \pm .132$]

\bottomrule
\end{tabular}
\end{adjustbox}
\end{table*}

\subsubsection{Cross-noise type generalisation}
This experiment investigates how the trained models generalise to unseen noise types. First we keep away one single noise type as unseen during training and mix it with test speech blocks. Then we use all of the rest of the noise types for training the model, and the training data with multiple noise types does share the same training speech blocks.

The validation set uses the same noise-types as training, but they have their own validation speech blocks. Note that there is still no overlap between the sample blocks in the training, validation and test sets. Due to space limitation, we show only \textit{bbl} as unknown. 
Then we also test the \textit{bbl} model on the rest of noise types and the clean model on all noise types. We train models using noisy signals mixed at all SNRs and test the models on the individual SNRs. In the end, we also compute the average performances across SNRs and seedings. 

The results are shown in Table \ref{tab:rocauc_generalisation}. It is noted that training on multiple noise types (widely trained) and testing on a single one outperforms the opposite and the model trained with clean data. 
Inverse filter bank features remain to be superior. 


\subsection{Cepstral feature projection}

\begin{figure}[!ht]
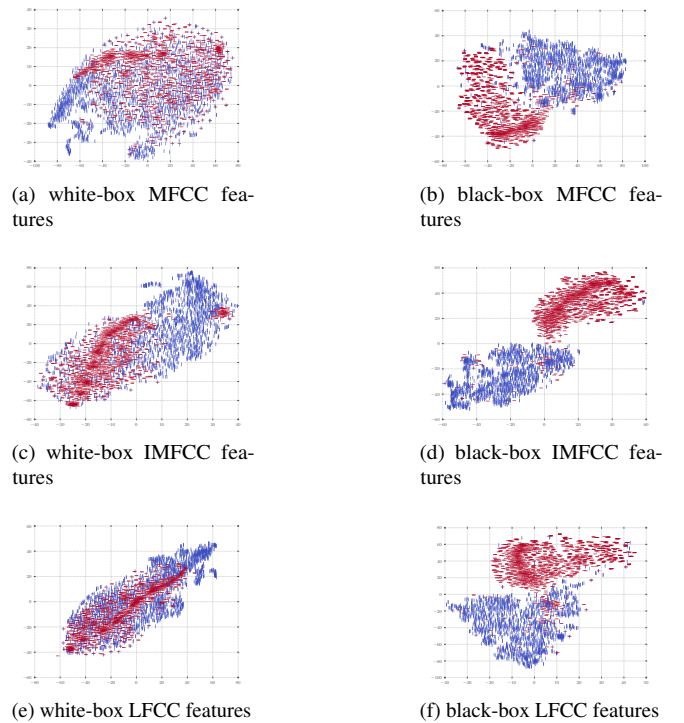

    \centering
    \foreach \transName in {MFCC,IMFCC,LFCC} {
    \subfloat[white-box \reprname{\transName} features\label{fig:tsne_proj_wba\transName}]{
        \includegraphics[width=3cm]{colorful/tsne/\transName_A}
      }
      \hfill
     \subfloat[black-box \reprname{\transName} features\label{fig:tsne_proj_wbb \transName}]{
        \includegraphics[width=3cm]{colorful/tsne/\transName_B}
      }
    }
    \caption{2D TSNE projection of $3500$ data points, blue |'s denotes benign data points while red -'s denotes adversarial data points. Left side is the white-box attacks and the right is the black-box attacks.}\label{fig:tsne_proj_wb1}
\end{figure}

We perform a projection of each supervector of cepstral features, from $31*20 = 620 \to 2$, to visualise if any low dimensional structure exists. We limit the number of random samples in the projection to $3500$ blocks, from the training subset of white- and black-box adversarial attack data sets. For the projection method we use T-distributed Stochastic Neighbour Embedding(TSNE) \cite{vanDerMaaten2008}. The results for MFCC, IMFCC and LFCC features are shown in Figure \ref{fig:tsne_proj_wb1}. Similar behaviour is as well observed for GFCC and IGFCC features, which is not shown due to space limit.  

Surprisingly although there is only a small discrepancy between the white-box and black-box attack classification performances, it is much easier to find dichotomic projections of the black-box attacks than those of white-box attacks, i.e. projected data of black-box attacks are much more separated from those of benign  than they are for white-box attacks. 

\section{Conclusion}
We investigated the effect of filter banks on detecting white-box and black-box adversarial attacks in original signals, signals with additive noise, speech only and non-speech only parts.
Our experiments show a clear tendency that the inverse filter bank based features outperform their counterparts and the linearly spaced filter bank based feature. Consistently through near to all trials we observe that inverse Mel-frequency cepstral coefficient and inverse Gammatone frequency cepstral coefficients based models are showing higher precision and recall measures than the rest of cepstral features spaces, indicating that in order to better detect adversarial attacks one should prefer the inverse filter banks over their counterparts. 

A similar observation is obtained when testing on noisy signals. One exception is the \textit{kitchen} noise type, which has highly flat long term average spectrum, and thus linear frequency cepstral feature performs the best. We also show that adversarial detection is effective in both speech only and non-speech only parts.


Future work includes investigating the use of learnt features for detecting adversarial examples, as learnt filter bank cepstral coefficients have shown to be highly effective for voice spoofing detection \cite{yu2017dnn}. 

\section{Acknowledgements}
We acknowledge the students Amalie V. Petersen, Jacob T. Lassen and Sebastian B. Schiøler at Aalborg University for conducting initial experiments.

 
\bibliography{biblio}

\end{document}